\documentclass[aps, prc, preprint, superscriptaddress, nofootinbib]{revtex4}

\usepackage{CJK}
\usepackage[usenames]{color}
\usepackage{graphicx}
\usepackage{amsmath}
\usepackage{amsfonts}
\usepackage{amssymb}
\usepackage{mathrsfs}
\usepackage{bm}
\usepackage{verbatim}
\usepackage{cancel}
\usepackage{enumitem}

\newcommand{\mhi}{M_{\text{hi}}}

\preprint{CTP-SCU/2017009}

\graphicspath{{../}{./plots/}}

\begin{document}

\begin{CJK*}{UTF8}{gbsn}
\title{Two bosons in a narrowly resonant trap}

\author{Bingwei Long (龙炳蔚)}
\affiliation{Center for Theoretical Physics, Department of Physics, Sichuan University, 29 Wang-Jiang Road, Chengdu, Sichuan 610064, China}

\author{Jianfeng Wang (王健峰)}
\affiliation{Center for Theoretical Physics, Department of Physics, Sichuan University, 29 Wang-Jiang Road, Chengdu, Sichuan 610064, China}

\author{Songlin Lyu (吕松林)}
\email{songlinlv@stu.scu.edu.cn}
\affiliation{Center for Theoretical Physics, Department of Physics, Sichuan University, 29 Wang-Jiang Road, Chengdu, Sichuan 610064, China}

\date{April 27, 2017}

\begin{abstract}
It is rare for an $S$-wave resonance to remain narrow while approaching threshold as a result of the interaction parameters being fine tuned. Such an interaction is, however, realized by pion-charmed baryon system $\pi \Sigma_c$, which has a narrow resonance $\Lambda_c^+(2595)$ located a few MeVs above its threshold. We study the consequence of narrowly resonant inter-species interaction in a three-body system composed of two identical light bosons and an extremely heavy particle. We concern ourselves in the present paper with resonances of the three-body system, as opposed to bound states. The focus is on the technical aspects of constructing the integral equation for three-body amplitudes and of analytically continuing the integral equation to locate the ``trimer'' resonance pole in the unphysical sheet of the center-of-mass energy plane.
\end{abstract}

\maketitle
\end{CJK*}

\section{Introduction\label{sec-intro}}

From the viewpoint of nonrelativistic potential models for single-channel processes, it is rather difficult to generate an $S$-wave resonance, because the potential, without assistance of centrifugal forces, will have to be tuned in a certain way so that a barrier will be in place to prevent the constituents from escaping. Even finer tuned is the case where the $S$-wave resonance sits very close to the two-body threshold and, in the mean time, remains a narrow profile. Remarkably, there exists an example of near-threshold $S$-wave resonances in hadronic physics: Charmed baryon $\Lambda_c^+(2595)$ with quantum numbers $I(J^P) = 0(\frac{1}{2}^-)$ can be thought of as the $S$-wave resonance located about $2$ MeV above the threshold of two-body continuum consisting of a pion and another charmed baryon $\Sigma_c$ with $I(J^P) = 1(\frac{1}{2}^+)$ ~\cite{Olive:2016xmw}. (For studies of similar or different opinions regarding $\Lambda_c^+(2595)$, see Refs.~\cite{Hyodo-2013iga, Long-2015pua, Hofmann, Mizutani, Romanets, Liang-2014kra, Lu-2014ina}.)

Using a general framework based on effective field theory (EFT), we discuss the implication of a narrowly resonant inter-species interaction for the three-body system composed of two identical light bosons and a heavy particle, with the heavy one being either bosonic or fermionic. Inspired by the physical example of $\pi \Sigma_c$ system, in which the masses of the pions ($\Sigma_c$'s) average out on isospin to be $138$ ($2455$) MeV, we consider in the first study the vanishing limit of the light-to-heavy mass ratio. We will see that there exists a trimer resonance near the three-body threshold. While that result is interesting in its own right (for example, it led to the prediction of a three-body hadronic state made up of $\pi \pi \Sigma_c$~\cite{Long-2016oog}), the main goal here is to explain the technique of analytically continuing the three-body amplitude in the complex plane of the center-of-mass (CM) energy, in search for the trimer resonance pole.

Suppose that the energy shift from the boson-heavy particle threshold to the dimer resonance peak is so small that momenta exchanged between the light boson and heavy particle are much smaller than the mass of the exchanged force-mediator particle. In such a case, an EFT with only contact interactions can be constructed, which has a priori the breakdown scale $\mhi$ about the mass of the exchanged particle~\cite{vanKolck-1997ut, VanKolck98, KSW98bis, KSW98}. In two-body scattering, this EFT is equivalent to the effective range expansion of the scattering amplitude. For the $S$ wave, the amplitude has the following form,
\begin{equation}
    f_{2 b} = \frac{1}{-1/a_0 + \frac{r_0}{2} k^2 - ik}\, ,
\end{equation}
where $k$ is the CM momentum, $a_0$ the scattering lenght and $r_0$ the effective range. The expansion implies two poles in the complex $k$ plane:
\begin{equation}
    k_\pm = \frac{1}{r_0} (i \pm \sqrt{\frac{2r_0}{a_0} - 1}) \, .
\end{equation}
In order for there to be a near-threshold, narrow resonance, both real and imaginary parts of $k_\pm$ must be very small $k_\pm \ll \mhi$ and $r_0 < 0$. This is achieved by having at least $|r_0| \gg \mhi^{-1}$. Without losing generality, we let $|r_0|/a_0$ be a free parameter. In the example of $\pi \Sigma_c$, the light quark masses as free parameters of quantum chromodynamics can be tuned to make $|r_0| \gg \mhi^{-1}$ and $|r_0|/a_0$ an adjustable parameter, as shown in Ref.~\cite{Long-2015pua}. The scenario $a_0 < 0$ and $|r_0/a_0| \gg 1$ is considered in Ref.~\cite{Bedaque-2003wa}.

Moreover, we assume that the light boson and heavy particle both have $N_f$ flavors and that they interact resonantly only through the flavor-singlet channel. Although this is configured so as to emulate the $\pi \Sigma_c$ system of which $\Lambda_c^+(2595)$ is an isoscalar resonance, the $N_f$ dependence may find relevance in, for instance, atomic systems in which the heavy particle is a bosonic atom with integer spin and it interacts with the light bosonic atom only through spin-singlet states.

The light bosons in the three-body system are assumed to interact so weakly with each other that the boson-boson interactions can be neglected. This assumption is made to retain an important feature of $\pi \pi \Sigma_c$, in which the pions are very soft, with three-momenta $Q \sim 20$ MeV around the resonance peak; therefore, their couplings are weak, proportional to $Q$ or $m_\pi^2$, because of their pseudo Goldstone-boson nature. Despite this presumption, the three-body system under consideration is not the textbook problem of a non-interacting bosonic system whose wave function factorizes into product of individual bosonic wave functions, because the flavor index of the heavy particle and energy dependence of the narrowly resonant potential make it impossible to separate bosonic coordinates in the three-body Schr\"odinger equation.

The theoretical framework is established in Sec.~\ref{sec-framework}, where the integral equation for the three-body amplitude is derived. Section~\ref{sec-cont} discusses the singularities of the integral equation and how the equation can analytically continued, with the numerical results shown toward the end. A summary is offered in Sec.~\ref{sec-summary}.

\section{Framework\label{sec-framework}}

The effective Lagrangian terms relevant for the paper include
\begin{equation}
\begin{split}
  \mathcal{L} &= \sum_a \phi_a^\dagger \left(i\partial_t + \frac{\nabla^2}{2m}\right)\phi_a + \sum_a i\psi^\dagger_a \dot{\psi_a} + \Delta^\dagger \left(i\partial_t - \delta \right) \Delta \\
  & \quad \quad + h \sqrt{\frac{2\pi}{mN_f}} \sum_a \left(\Delta^\dagger \psi_a \phi_a + h.c. \right) + \cdots \, ,
\end{split}
\end{equation}
where $\phi_a$ denotes the field that annihilates a light boson with mass $m$ and flavor index $a$, $\psi_a$ corresponds to the heavy particle with flavor index $a$, and $\Delta$ is the flavor-singlet field responsible for the dimer resonance with an energy above the $\phi \psi$ threshold by $\delta$. The ``...'' serves as a reminder that other symmetry-observing terms can be accounted for as subleading corrections. As already mentioned, we consider the limit where the light-to-heavy mass ratio vanishes. This is reflected by the absence of the kinetic term $\nabla^2/2m_\psi$ ($\nabla^2/2m_\Delta$) for $\psi_a$ ($\Delta$), that is, $\psi_a$ or $\Delta$ does not propagate in space as long as only one heavy particle is present. Moreover, it does not affect the discussion whether $\psi_a$ is fermion or boson as far as the present paper is concerned. This sort of frameworks for few-body dynamics were also found in Refs. \cite{Kaplan-1996nv, Bedaque-1998km, Bedaque-1998kg} that use auxiliary fields to construct resonant two-body interactions.

\begin{figure}
    \centering
    \includegraphics[scale=1.0]{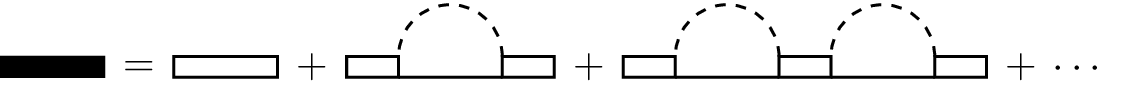}
    \caption{Dressing the $\Delta$ propagator. The solid (dashed) line is the propagator of $\psi$ ($\phi$), and the double line represents propagation of $\Delta$.}
    \label{fig_dressed_lyu}
\end{figure}

\begin{figure}
    \centering
    \includegraphics[scale=0.8]{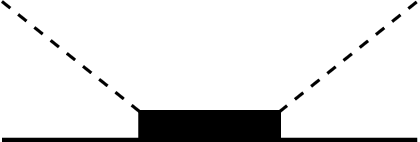}
    \caption{$\phi \psi$ elastic scattering.}
    \label{fig_scattering_lyu}
\end{figure}

The inter-species interaction between $\psi_a$ and $\phi_a$ is encapsulated in the dressed $\Delta$ propagator, diagrammatically presented in Fig.~\ref{fig_dressed_lyu},
\begin{equation}
    i D(p) = \frac{i}{p_0 - \delta - h^2\sqrt{- 2 p_0 m -i0} + i0} \, ,
\end{equation}
where $p \equiv (p_0, \vec{p}\,)$ denotes the energy and momentum flowing through. Figure~\ref{fig_scattering_lyu} shows that the amplitude for singlet-channel $\phi \psi$ elastic scattering is readily obtained by attaching the transition vertex of $\Delta \to \phi \psi$ to the dressed propagator:
\begin{equation}
    T_{\phi \psi} = \frac{4\pi}{-1/a_0 + \frac{r_0}{2} k^2 - ik}\, ,
\end{equation}
where
\begin{equation}
    1/a_0 = - \frac{\delta}{h^2} \quad  \text{and} \quad r_0 = -\frac{1}{h^2 m} \, .
\end{equation}

The $\phi \phi \psi$ system underpinned by the flavor-singlet interaction can be accessed by correlation functions of the following type,
\begin{equation}
\sum_{b b'}\langle 0 | T\{ \psi_b \phi_b \phi_a  \phi_a^\dagger \psi_{b'}^\dagger \phi_{b'}^\dagger) \} |0\rangle \, .
\end{equation}
We do not need all possible information about the dynamics in this channel but the trimer pole position. One of the more convenient vehicles to probe that is the $\phi_a \Delta$ amplitude, more precisely, the amputated correlation function of the form
\begin{equation}
    \begin{split}
    &\int d^4 x' d^4 y' d^4 x d^4 y\, e^{i (q\cdot x' + p'\cdot y' - k\cdot x - p\cdot y)} \langle 0|T\{ \Delta(x') \phi_{a'}(y') \phi_a^\dagger(y) \Delta^\dagger(x) \}|0 \rangle \, , \\
    &\quad \equiv (2\pi)^4 \delta^{(4)}(p + k - p' - q) \delta_{a' a} \mathcal{M}(\vec{k}, \vec{q}; E, B, q_0)
    \end{split}
\end{equation}
where $k = (k_0, \vec{k}\,) $ [$p = (-B, -\vec{k})$] is the incoming four-momentum of $\phi_a$ ($\Delta$) in the CM frame and $q = (q_0, \vec{q}\,)$ [$p' = (E-q_0, -\vec{q}\,)$] is the outgoing four-momentum. We wish to determine the pole position of $\mathcal{M}$ as a function of the CM energy $E$ while other kinematic variables are fixed.

\begin{figure}
    \centering
    \includegraphics[scale=0.6]{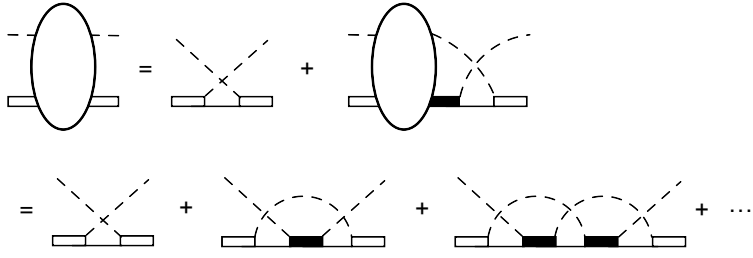}
    \caption{$\phi \Delta$ amplitude.}
    \label{fig_blob}
\end{figure}

With the boson-boson interaction negligible, any possible nontrivial formation of $\phi  \phi \psi$ must be a result of two bosons coming alternatively to interact with the heavy particle. Figure~\ref{fig_blob} shows the structure of the $\phi \Delta$ amplitude. The first line is diagrammatic representation of the integral equation for the (off-shell) amplitude, which, upon expansion, is resummation of $s$-channel exchanges of $\psi$ shown in the second line. The interacting $\phi \psi$ subsystem buried in the dressed $\Delta$ propagators, together with a second $\phi$ line floating around, demonstrates propagation of the $\phi \phi \psi$ states.

The integral equation is read off from the diagram as
\begin{equation}
\begin{split}
    \mathcal{M}(q_0; E, B) &= \frac{2\pi h^2}{m N_f} \frac{1}{B + q_0 - i0} + i \frac{2\pi h^2}{mN_f} \int \frac{d^4 l}{(2\pi)^4} \left[ \frac{1}{E - q_0 - l_0 + i0} \right. \\
    &\qquad\qquad \left. \times \frac{1}{l_0 - l^2/2m + i0} \frac{\mathcal{M}(l_0; E, B)}{E - l_0 - \delta - h^2\sqrt{-2m(E - l_0) -i0} + i0} \right] \, , \label{eqn_Md4l}
\end{split}
\end{equation}
where $l \equiv |\vec{l}|$. Because the driving term and propagators of the integrand do not explicitly depend on $\vec{k}$ or $\vec{q}$, $\mathcal{M}$ do not depend explicitly on them either.

Before discussing how to analytically continue the amplitude, we assume $E$, $B$, and $q_0$ to take real values. $\mathcal{M}(q_0; E, B)$ as a function of $q_0$ has an upper-half-plane pole contributed by the inhomogeneous term on the right hand side of Eq.~\eqref{eqn_Md4l}, but $\mathcal{M}(q_0; E, B)$ has no pole in the lower-half plane, a claim which we will come back to in next paragraph. Therefore, we can integrate over $l_0$ by enclosing the contour in the lower-half $l_0$ plane and picking up the pole of the first propagator in the second line of Eq.~\eqref{eqn_Md4l}, $l_0 = l^2/2m-i0$, and then we arrive at
\begin{equation}
\begin{split}
    \mathcal{M}(q_0; E, B) &= \frac{2\pi h^2}{m N_f} \frac{1}{B + q_0 - i0} + \frac{4\pi h^2}{N_f} \int \frac{d^3 l}{(2\pi)^3} \left[ \frac{1}{2m(E - q_0) - l^2 + i0} \right. \\
    &\qquad\qquad \left. \times \frac{2m\,\mathcal{M}(l^2/2m; E, B)}{2m\left(E - \delta - h^2\sqrt{l^2 -2mE -i0}\right) - l^2 + i0} \right] \, . \label{eqn_Md3l}
\end{split}
\end{equation}
It can be verified that the remaining integral does not contribute any $q_0$ poles to $\mathcal{M}(q_0; E, B)$, but a branch cut starting at $q_0 = E$ owing to the first propagator inside the integral, an instance of so-called endpoint singularity~\cite{Eden-Smatrix}.

Had $\mathcal{M}(q_0; E, B)$ had an isolated pole in the lower-half $q_0$ plane at, say, $q_0 = \mathcal{Z}_i$,
\begin{equation}
\mathcal{M}(q_0; E, B) \to \frac{R_i(E, B)}{q_0 - \mathcal{Z}_i}\quad \text{when}\; q_0 \to \mathcal{Z}_i\, \; \left(\text{Im} \mathcal{Z}_i < 0 \right)\, ,
 \end{equation}
it would have added to the right hand side of Eq.~\eqref{eqn_Md3l} the following three-dimensional integral,
\begin{equation}
    \frac{4\pi h^2}{N_f} \int \frac{d^3 l}{(2\pi)^3} \frac{1}{E - q_0 - \mathcal{Z}_i} \frac{1}{2m\mathcal{Z}_i-\vec{l}^2} \frac{R_i(E, B)}{E - \mathcal{Z}_i - \delta -h^2 \sqrt{2m(\mathcal{Z}_i - E)}} \, .
\end{equation}
The above integral could have only contributed another cut in the $q_0$ plane, again, an endpoint singularity. This, however, would contradict the assumption of $\mathcal{M}(q_0; E, B)$ having poles in the lower-half $q_0$ plane.

We can simplify further Eq.~\eqref{eqn_Md3l} by integrating over the angular parts of $\vec{l}$, trading $h^2$ and $\delta$ for two-body scattering parameters $a_0$ and $r_0$, and setting $q_0 = q^2/2m$:
\begin{equation}
\begin{split}
    & t(q; \mathcal{E}, \mathcal{B}) = \frac{8\pi/|r_0|}{N_f (q^2 + \mathcal{B})} + \frac{2}{\pi N_f} \int^\Lambda dl \frac{l^2}{q^2 - \mathcal{E} + l^2 + i0} \\
    & \quad \times \frac{t(l; \mathcal{E}, \mathcal{B})}{-\frac{1}{a_0} - \frac{|r_0|}{2}(\mathcal{E} - l^2) + \sqrt{l^2 - \mathcal{E} - i0} -i0} \, , \label{eqn_1dinteqn}
\end{split}
\end{equation}
where $t(q; \mathcal{E}, \mathcal{B}) \equiv 2m \mathcal{M}(q^2/2m; E, B)$ with $\mathcal{E} \equiv 2m E$ and $\mathcal{B} \equiv 2 m B$. An ultraviolet cutoff $\Lambda$ is put in place to remind us that regularization is in principle needed to have the integral equation well defined.

We would like any conclusions derived from Eq.~\eqref{eqn_1dinteqn} to be as insensitive as possible to the \emph{detail} of short-range physics that is not manifest in the framework. To avoid excessive modeling of short-range physics in any EFT, one must require as a necessity any observable be independent of the cutoff parameter $\Lambda$ that is introduced to regularize loop integrals. This could constrain in surprisingly non-trivial ways the interaction terms to be included at a given order of EFTs.

For example, in the EFT formulation of three identical bosons systems where the two-body scattering length diverges ~\cite{Bedaque-1998km, Bedaque:2000ft, Barford:2002ut, Platter:2006ev}, the instability of the three-body amplitude against variation of $\Lambda$ demands the three-body force be added at leading order (LO) so that its running can absorb the cutoff dependence of three-body observables. The lesson for the problem in hand is the following: If the integral in Eq.~\eqref{eqn_1dinteqn} is convergent in the sense that its solution stabilizes when the cutoff $\Lambda \to \infty$, no information about three-body forces is required at LO and Eq.~\eqref{eqn_1dinteqn} will be self-consistent.

\begin{figure}
    \centering
    \includegraphics[scale=0.6]{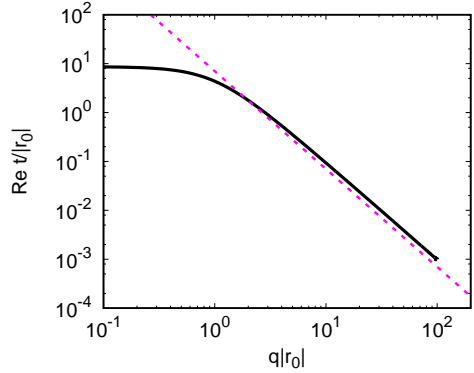}
    \caption{Represented by the solid line is $\text{Re}\, t(q; \mathcal{E}, \mathcal{B})$ as a function of $q$, and the dashed line $\propto 1/q^2$. See the text for the parameters used to plot $\text{Re}\, t(q; \mathcal{E}, \mathcal{B})$.}
    \label{fig_tvsq}
\end{figure}

Equation~\eqref{eqn_1dinteqn} indeed converges. Plotted in Fig.~\eqref{fig_tvsq} is $\text{Re}\, t(q; \mathcal{E}, \mathcal{B})/|r_0|$ as a function of $q|r_0|$ for the following values taken by the parameters:
\begin{equation}
    \frac{|r_0|}{a_0} = -2\, , \quad \mathcal{E} |r_0|^2 = -0.5 + 0.01i\, , \quad \mathcal{B} |r_0|^2 = 1.0 - 0.01 i\, , \quad \Lambda |r_0| = 100 \, .
\end{equation}
We have varied the value of $\Lambda|r_0|$ from $100$ to $20$, and the resolution of Fig.~\ref{fig_tvsq} is unable to show the variation of $t(q; \mathcal{E}, \mathcal{B})$ up to the range of $q$ allowed by $\Lambda$. Perhaps more pertinently, Fig.~\ref{fig_tvsq} indicates the asymptotic behavior of $t(q; \mathcal{E}, \mathcal{B})$ when $q^2 \gg \mathcal{E}$, where $t(q; \mathcal{E}, \mathcal{B})$ is dominated by its real part. Assisted by a ``ruler'', the dashed line representing a function $\propto 1/q^2$ in the log-log plot, we can see that $t(q; \mathcal{E}, \mathcal{B})$ $\propto 1/q^2$ in far off-shell region, in agreement with the naive expectation of the $q$-dependence from the right hand side of Eq.~\eqref{eqn_1dinteqn}. With such a suppression of $t(l; \mathcal{E}, \mathcal{B})$, the integrand in Eq.~\eqref{eqn_1dinteqn} goes as $1/l^4$ when $l$ is large. Therefore, the integral is expected to converge.

\section{Analytic continuation\label{sec-cont}}

The main technical challenge of the work is to continue analytically Eq.~\eqref{eqn_1dinteqn} into the complex $\mathcal{E}$ plane. The key is to deform tactfully the integration contour so that as $\mathcal{E}$ moves into its unphysical sheet, the contour does not interfere singularities of the integrand, including those of both propagators and $t(l; \mathcal{E}, \mathcal{B})$ itself as a function of $l$. We need to let $q$ be defined along the same contour as $l$, so that the integral equation can, upon discretization, turn into a linear system of equations that could be solved numerically.

A general methodology concerning analytic continuation of three-body integral equations is discussed in Ref.~\cite{Pearce-1984ca}, but not accounting for the singularities of the undetermined three-body amplitude itself [in our case, $t(l; \mathcal{E}, \mathcal{B})$] makes its analysis less rigorous than what is presented here.

It proves convenient to illustrate the contour, instead of $l$, by $\omega \equiv \mathcal{E} - l^2$, which is the energy flowing through the dressed $\Delta$ propagator. Once the analytic structure is understood, we will switch back to the $l$ plane in numerical calculations after the subtle difference between the integrations over $\omega$ and $l$ is explained.

\begin{figure}
    \centering
    \includegraphics[scale=1.0]{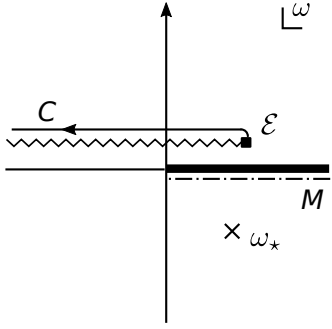}
    \caption{The integration contour $C$ and singularities of the integrand in the $\omega$ plane when $\mathcal{E}$ is in the physical sheet. The thick line corresponds to singularity \ref{sqrtcut}, the dot-dashed line $M$ singularities \ref{poleline} and \ref{tmatrix_cut}, the cross singularity \ref{polestar}, and the zigzag singularity \ref{zigzag}.}
    \label{fig_1stsheet}
\end{figure}

Replacing $l$ with $\omega$, we rewrite Eq.~\eqref{eqn_1dinteqn},
\begin{equation}
\begin{split}
t(\omega_q; \mathcal{E}, \mathcal{B}) = \frac{8\pi/|r_0|}{N_f (\mathcal{E}+\mathcal{B} - \omega_q)} + \frac{1}{\pi N_f} \int_C d\omega\, \frac{\sqrt{\mathcal{E} - \omega}}{\mathcal{E} - \omega_q - \omega} \frac{t(\omega; \mathcal{E}, \mathcal{B})}{-\frac{1}{a_0} - \frac{|r_0|}{2} \omega + \sqrt{-\omega}} \, , \label{eqn_omegainteqn}
\end{split}
\end{equation}
where $\omega_q \equiv \mathcal{E} - q^2$.
Shown in Fig.~\ref{fig_1stsheet}, the contour $C$ in the $\omega$ plane starts at $\mathcal{E}$ and extends leftward to infinity, when $\mathcal{E}$ is in the physical sheet and it approaches the real axis from above. Let us take stock of singularities of the integrand as a function of $\omega$ and illustrate them in Fig.~\ref{fig_1stsheet}:
\begin{enumerate}[label=(\roman*)]
    \item \label{sqrtcut} The origin is a branch point of $\sqrt{-\omega}$, with the cut represented by the thick line in Fig.~\ref{fig_1stsheet} that lies along the positive real axis, where the two Riemann sheets of $\sqrt{-\omega}$ meet.

    \item \label{poleline} $t(\omega; \mathcal{E}, \mathcal{B})$ is analytic everywhere on $C$, so must be $t(\omega_q; \mathcal{E}, \mathcal{B})$. Therefore, for each and every $\omega_q$ on $C$, the $\omega$ pole of the boson propagator $(\mathcal{E} - \omega - \omega_q)^{-1}$,
    \begin{equation}
      \omega^\text{pole} = \mathcal{E} - \omega_q \, , \label{eqn-pole-cprime}
    \end{equation}
    must not lie on $C$. As $\omega_q$ starts from $\mathcal{E}$ and moves along $C$, the corresponding $\omega^\text{pole}$ moves too, depicting a trajectory that is illustrated by the dot-dashed line $M$ in Fig.~\ref{fig_1stsheet}. The condition that $\omega^\text{pole}$, for any $\omega_q$, must not lie on $C$ is equivalent to the one that $C$ and $M$ must not intersect. (Note that $C$ and $M$ are symmetric about $\omega = \mathcal{E}/2$.)

    \item \label{polestar} The dimer resonance pole of the dressed $\Delta$ propagator, $\omega_\star(a_0, r_0)$, is represented by the cross in the second sheet of $\sqrt{-\omega}$. There exists actually a second pole, but it is of much less concern for the region of $\mathcal{E}$ to be explored in the paper.

    \item \label{zigzag} The branch cut of $\sqrt{\mathcal{E} - \omega}$ in the numerator is represented by the zigzag in Fig.~\ref{fig_1stsheet}.

    \item Let us turn to the $\omega$ singularities of $t(\omega; \mathcal{E}, \mathcal{B})$ itself. Once a legal contour is chosen, the right hand side of Eq.~\eqref{eqn_1dinteqn} gives a representation for $t(\omega_q; \mathcal{E}, \mathcal{B})$, even for $\omega_q$ that is not necessarily on $C$. The driving term $\propto (\mathcal{E} + \mathcal{B} - \omega_q)^{-1}$ contributes to $t(\omega_q; \mathcal{E}, \mathcal{B})$ a pole at $\omega_q = \mathcal{E} + \mathcal{B}$. However, it is of little concern to us because we are free to choose any value for $\mathcal{B}$, provided that the trimer pole of $t(\omega_q; \mathcal{E}, \mathcal{B})$ as a function of $\mathcal{E}$ is all we care for.

    \item \label{tmatrix_cut} The integration in Eq.~\eqref{eqn_1dinteqn} contributes a branch cut to $t(\omega_q; \mathcal{E}, \mathcal{B})$ as a function of $\omega_q$. We first notice that $\omega_q = 0$ is an endpoint singularity~\cite{Eden-Smatrix}, for no matter how $C$ is deformed the denominator of $(\mathcal{E} - \omega - \omega_q)^{-1}$ will necessarily vanish for $\omega_q \to 0$ at $\omega = \mathcal{E}$, where the contour starts. The associated cut line is defined by
    \begin{equation}
      \omega_q^{\text{cut}} = \mathcal{E} - \omega \, .\label{eqn-cut-cprime}
    \end{equation}
    As $\omega$ moves along $C$, the trajectory of $\omega_q^{\text{cut}}$ is precisely $M$. Therefore, if $C$ and $M$ do not intersect as \ref{poleline} already requires, the branch cut of $t(\omega; \mathcal{E}, \mathcal{B})$ as a function of $\omega$ is automatically avoided when $\omega$ is integrated over.

\end{enumerate}

\begin{figure}[tb]
    \centering
    \includegraphics[scale=1.1]{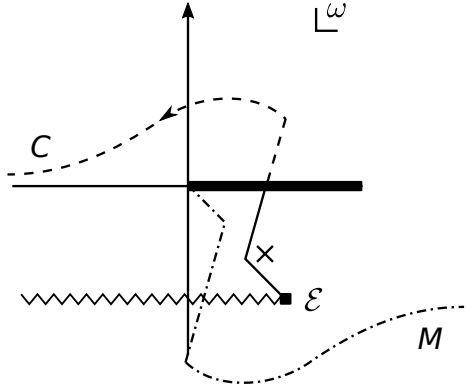}
    \caption{Deforming the contour $C$ when $\mathcal{E}$ is in the unphysical sheet. The label $\omega_\star$ on the cross is omitted to avoid congestion of symbols. See the text for detailed explanation.}
    \label{fig_2ndsheet}
\end{figure}

As $\mathcal{E}$ moves downward, it will cross the branch cut~\ref{sqrtcut} if $\mathcal{E}$ is forced to stay in the physical sheet. As a result, the integration encounters a discontinuity if we do not deform $C$. This discontinuity gives rise to the three-body unitarity cut of $t(\omega_q; \mathcal{E}, \mathcal{B})$ as a function of $\mathcal{E}$. The unphysical sheet of $\mathcal{E}$ can be defined if $C$ deforms continuously in such a way that it does not cross or intersect any singularities listed above. Figure~\ref{fig_2ndsheet} illustrates one way to achieve this.

First, the starting point of $C$ ($\omega = \mathcal{E}$, denoted by the filled square) needs burrow through the second Riemann sheet of $\sqrt{-\omega}$ while the part of $C$ that is toward $-\infty$ stays in the first sheet (shown as the dashed line in Fig.~\ref{fig_2ndsheet}). $C$ must go about the origin counterclockwise, not only to avoid crossing the branch point of $\sqrt{-\omega}$ but also to prevent $C$ and $M$ from intersecting each other.

Second, $C$ must not cross through $\omega_\star$ (the cross in Fig.~\ref{fig_2ndsheet}). There are two ways for the contour to circumvent $\omega_\star$: clockwise or counterclockwise. The difference between the two choices amounts to one of the discontinuities of $t(\omega_q; \mathcal{E}, \mathcal{B})$ as a function of $\mathcal{E}$, owing to another endpoint singularity of the integral: When $\mathcal{E} \to \omega_\star$, the contour can in no way avoid $\omega_\star$ by deformation, for it starts at $\omega_\star$. This branch point of $t(\omega_q; \mathcal{E}, \mathcal{B})$ as a function of $\mathcal{E}$ marks the boson-dimer resonance threshold. The associated cut line is conventionally chosen to lie horizontally to $+\infty$, and in accordance with that choice we let $C$ circumvent counterclockwise $\omega_\star$ like shown in Fig.~\ref{fig_2ndsheet}. However, at the cost of adding a counterclockwise infinitesimal circle around $\omega_\star$, we could let the contour go the other way instead, if it is more convenient in practice to do so.

\begin{figure}
    \centering
    \includegraphics[scale=0.6]{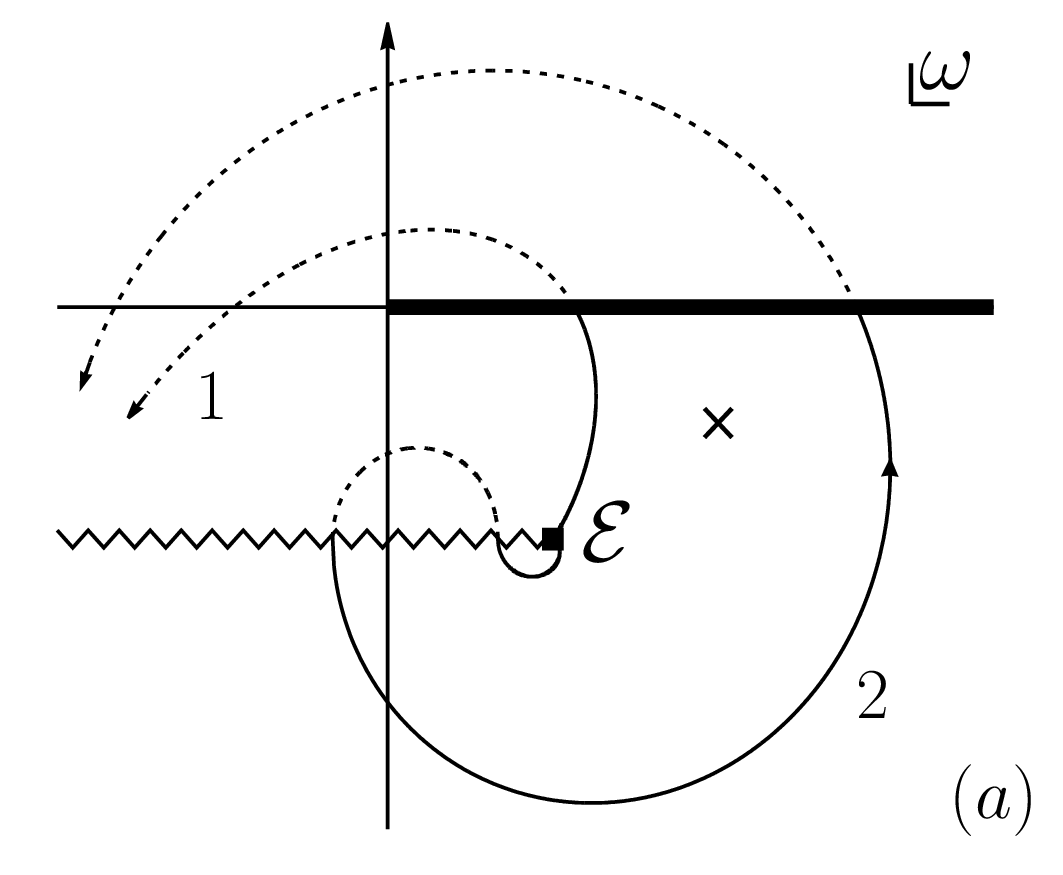}
    \includegraphics[scale=0.6]{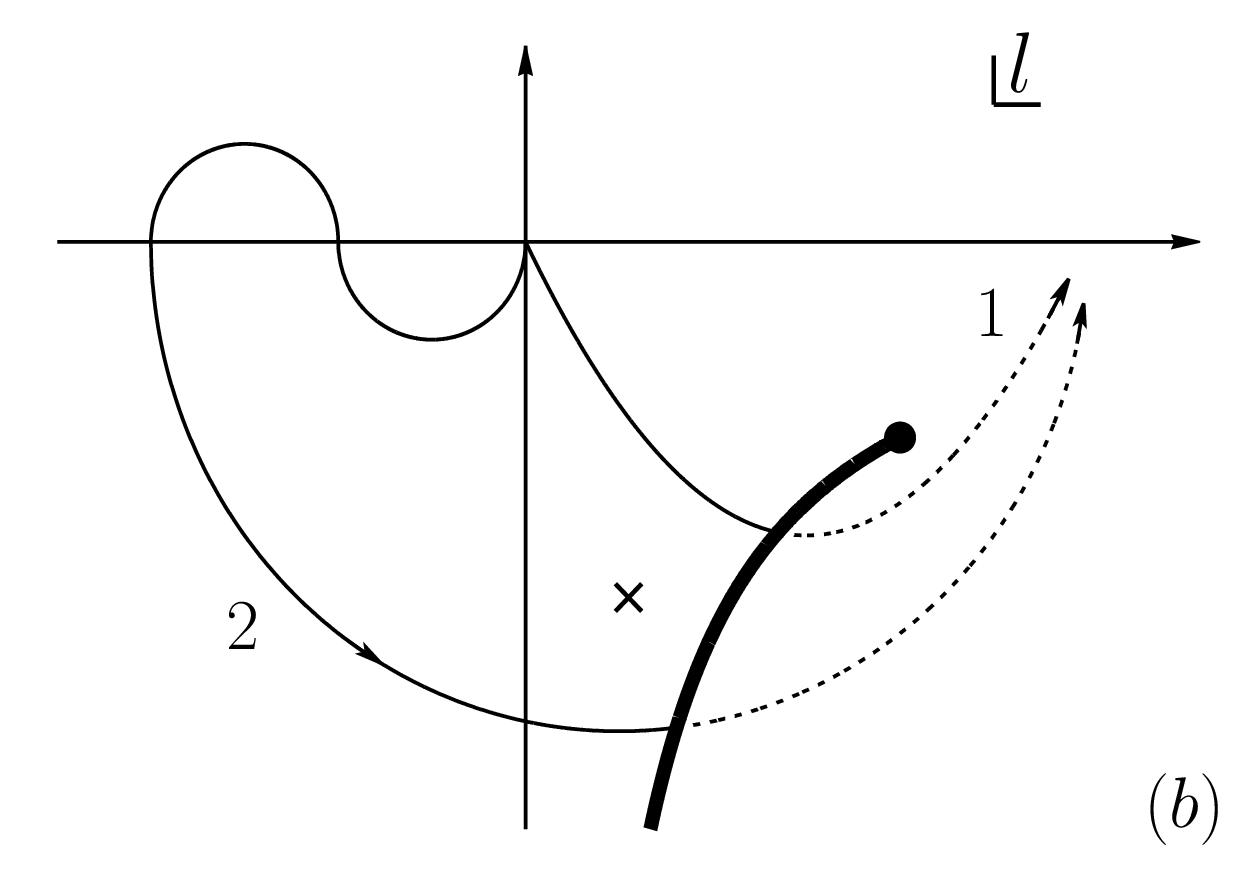}
    \caption{The mapping between $\omega$ and $l$. In the $\omega$ plane, the symbols are the same as in Fig.~\ref{fig_2ndsheet}. In the $l$ plane, the thick line is the branch cut of $\sqrt{l^2 - \mathcal{E}}$, which maps back onto the thick line in the $\omega$ plane.}
    \label{fig_mappinglyu}
\end{figure}

The numerical calculation is easier to implement with $l$ being the integration variable. The mapping
\begin{equation}
    l = \sqrt{\mathcal{E} - \omega}
\end{equation}
has a branch cut, which is just the zigzag in Figs.~\ref{fig_2ndsheet} and \ref{fig_mappinglyu}. It is customary to map different Riemann sheets of $\omega$ (due to the cut of $\sqrt{\mathcal{E} - \omega}$) separately onto upper and lower halves of the $l$ plane. Figure~\ref{fig_mappinglyu} demonstrates the mapping by showing two contours in both $\omega$ and $l$ planes. In the $\omega$ plane, only the second Riemann sheet is supposed to be visible, so the parts of the contours that burrow through other sheets are hidden, represented by the dashed lines. Particularly telling is the semicircle of contour $2$ lying in the upper half $l$ plane; it shows up in the second sheet of $\sqrt{\mathcal{E} - \omega}$ in the $\omega$ plane.

$\sqrt{-\omega}$ adds another branch cut to the $\omega$ plane, which maps onto the cut line in the $l$ plane due to $\sqrt{l^2 - \mathcal{E}}$ [in the integrand of Eq.~\eqref{eqn_1dinteqn}], both represented by thick lines in corresponding planes.

Contour $1$ satisfies the conditions laid out for analytic continuation. Contour $2$ is not equivalent to $1$ because they enclose $\omega_\star$ (in the $\omega$ plane), one of the poles of the integrand in Eq.~\eqref{eqn_omegainteqn}. But we can easily take it out by subtracting the residue at $\omega_\star$.

\begin{figure}
    \centering
    \includegraphics[scale=0.6]{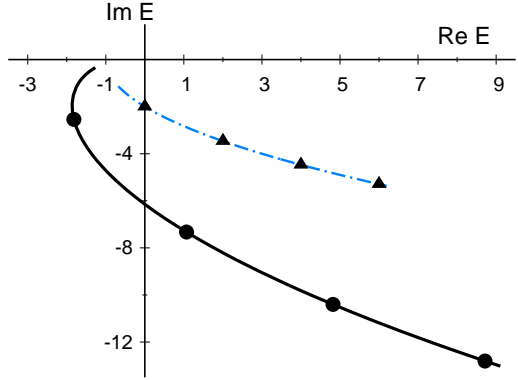}
    \caption{The solid line is the pole trajectory of the trimer resonance $E_{3b}$ as $|r_0|/a_0$ varies. The dot-dashed line is the threshold of boson-dimer resonance $E_{\phi \Delta}$. The unit of $E_{3b}$ and $E_{\phi \Delta}$ is $(2mr_0^2)^{-1}$. From right to left, the solid points and triangles correspond respectively to $|r_0|/a_0 = -4$,  $-3$, $-2$, and $-1$.}
    \label{fig_poletrajectory}
\end{figure}

\begin{figure}
    \centering
    \includegraphics[scale=0.7]{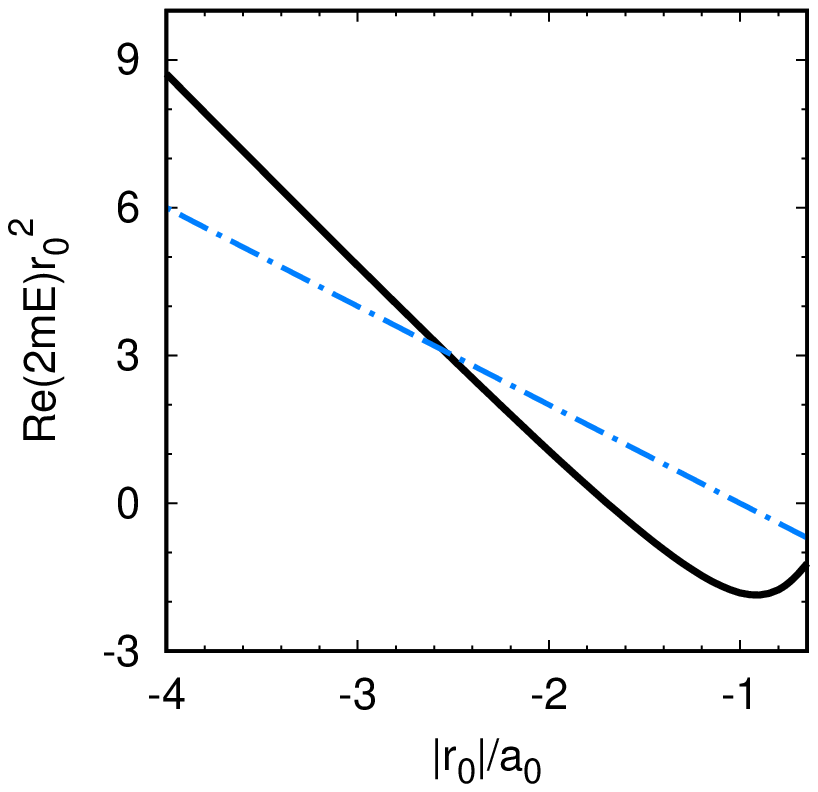}
    \hspace{5mm}
    \includegraphics[scale=0.7]{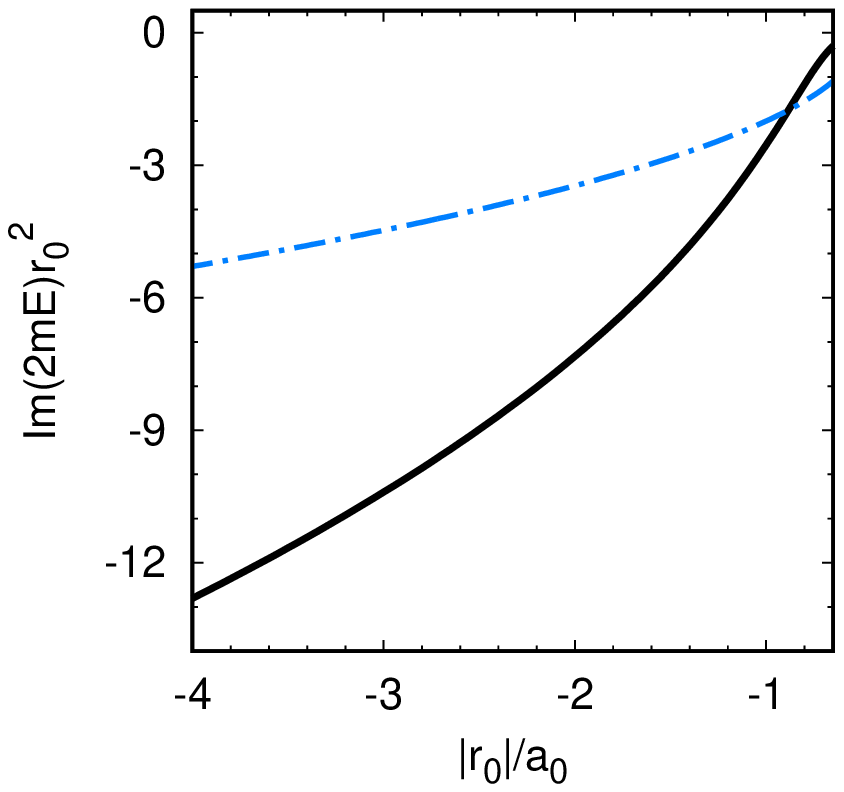}
    \caption{The solid line represents pole position of the trimer resonance as a function of $|r_0|/a_0$, with real and imaginary parts plotted in separate panels. The dot-dashed line is the threshold of boson-dimer continuum.}
    \label{fig_polevsa}
\end{figure}

It may be instructive to compare the trimer resonance pole to the threshold of boson-dimer $E_{\phi \Delta}$, which is always above the $\phi \phi \psi$ threshold by $\omega_\star/2m$: $E_{\phi \Delta} = \omega_\star/2m$. The trimer resonance pole is denoted by $E_{3b}$, the CM energy above the $\phi \phi \psi$ threshold. Dimensionless quantities $E_{3b} (2mr_0^2)$ and $E_{\phi \Delta} (2mr_0^2)$ depend only on the ratio $|r_0|/a_0$ and $N_f$. We use $N_f = 3$, in accordance with $\pi \pi \Sigma_c$ system that motivated this study, in numerical calculations to demonstrate the results. Figure~\eqref{fig_poletrajectory} shows the trajectories of $E_{3b}$ and $E_{\phi \Delta}$, as $|r_0|/a_0$ varies from $-4.1$ to $-0.65$, and Fig.~\ref{fig_polevsa} shows more explicitly how they change against $|r_0|/a_0$, with appropriate units.

As the magnitude of $|r_0|/a_0$ decreases, the inter-species attraction becomes stronger while the trimer resonance and the boson-dimer threshold both move closer to the real axis. But $E_{3b}$ approaches the $\phi \phi \psi$ threshold more rapidly than $E_{\phi \Delta}$ does. This is more quantitatively demonstrated in Fig.~\ref{fig_polevsa}. The real part of the trimer pole is seen to move below that of the boson-dimer threshold for $|r_0|/a_0 \gtrapprox -2.55$. The plot of imaginary parts suggests that when $|r_0|/a_0$ increases the trimer resonance becomes as stable as or even more so than the dimer resonance. Even though the trimer or the dimer is not actual bound state, a Borromean-like scenario is seen to emerge, in which the three-body system appears to posses more attraction than the stand-alone two-body system.

\section{Summary\label{sec-summary}}

We have considered the three-body system of two identical light bosons trapped by an extremely heavy particle through an inter-species interaction that generates a narrow, near-threshold $S$-wave resonance. The binary interaction is in fact realized in hadronic physics by $\pi \Sigma_c$, which has baryon resonance $\Lambda_c^+(2595)$ situated near its threshold. For a closer resemblance to $\pi \Sigma_c$ system, we have assigned $N_f$ flavors to both the bosons and the heavy particle, and have assumed that only the flavor-singlet channel is strong enough to induce a dimer resonance.

Much of the work is devoted to developing a machinery to analytically continue the three-body amplitude into the unphysical sheet of the center-of-mass energy $E_{3b}$, so that the trimer resonance pole can be extracted. The contour in the three-body integral equation was deformed so that the unphysical sheet of $E_{3b}$ can be accessed. We presented a detailed study of the singularities of the integral equation and showed how they can be circumvented in continuation.

The ultraviolet convergence of the integral equation indicates that the model dependence on short-range physics that is truncated from the low-energy theory is under control. On the other hand this implies that the three-body system does not exhibit Efimov states. This is actually consistent with the findings of Ref.~\cite{Helfrich-2010yr}, which showed that Efimov effect becomes less visible as the light-to-heavy mass ratio decreases.

With $(2mr_0^2)^{-1}$ as the unit, the trimer resonance pole $E_{3b}$ is a function of the ratio of the effective range to the scattering length $|r_0|/a_0$ and the number of flavors $N_f$. We showed for $N_f = 3$ how the trimer resonance pole moves as $|r_0|/a_0$ varies. When $|r_0|/a_0$ decreases to certain critical values the trimer resonance moves below the threshold of boson-dimer resonance, or it outlives the dimer resonance. This can be interpreted as the three-body system having more attraction than the inter-species two-body system. It may be interesting to study whether this feature persists for other values of $N_f$.

\acknowledgments

BwL thanks the organizers of the workshop ``Bound states in QCD and beyond II'' in St. Goar for the opportunity of communicating this work to broader audience. The work was supported in part by the National Natural Science Foundation of China (NSFC) under grant number 11375120.


\begin{thebibliography}{99}

\bibitem{Olive:2016xmw}
  C.~Patrignani {\it et al.} [Particle Data Group],
  Chin.\ Phys.\ C {\bf 40} (2016) no.10,  100001.

\bibitem{Hyodo-2013iga}
T.~Hyodo,
  Phys.\ Rev.\ Lett.\  {\bf 111} (2013) 132002.

\bibitem{Long-2015pua}
  B.~Long,
  Phys.\ Rev.\ D {\bf 94}, no. 1, 011503 (2016).

\bibitem{Hofmann}
J.~Hofmann and M.~F.~M.~Lutz,
Nucl.\ Phys.\ A {\bf 763}, 90 (2005).

\bibitem{Mizutani}
T.~Mizutani and A.~Ramos,
Phys.\ Rev.\ C {\bf 74}, 065201 (2006).

\bibitem{Romanets}
O.~Romanets, L.~Tolos, C.~Garcia-Recio, J.~Nieves, L.~L.~Salcedo, and
R.~G.~E.~Timmermans,
Phys.\ Rev.\ D {\bf 85}, 114032 (2012).

\bibitem{Liang-2014kra}
  W.~H.~Liang, T.~Uchino, C.~W.~Xiao, and E.~Oset,
  Eur.\ Phys.\ J.\ A {\bf 51} (2015) no.2,  16.

\bibitem{Lu-2014ina}
  J.~X.~Lu, Y.~Zhou, H.~X.~Chen, J.~J.~Xie, and L.~S.~Geng,
  Phys.\ Rev.\ D {\bf 92} (2015) no.1,  014036.


\bibitem{Long-2016oog}
  B.~Long,
  arXiv:1609.08940 [nucl-th].

\bibitem{vanKolck-1997ut}
  U. van Kolck,
  Lect. Notes Phys. {\bf 513} (1998) 62.

\bibitem{VanKolck98}
U. van Kolck,
Nucl. Phys. A {\bf 645} (1999) 273.

\bibitem{KSW98bis}
D.B.~Kaplan, M.J.~Savage, and M.B.~Wise,
Phys.\ Lett. B {\bf 424} (1998) 390.

\bibitem{KSW98}
D.B.~Kaplan, M.J.~Savage, and M.B.~Wise,
Nucl.\ Phys. B {\bf 534} (1998) 329.


\bibitem{Bedaque-2003wa}
  P.~F.~Bedaque, H.~W.~Hammer, and U.~van Kolck,
  Phys.\ Lett.\ B {\bf 569}, 159 (2003).

\bibitem{Kaplan-1996nv}
  D.~B.~Kaplan,
  Nucl.\ Phys.\ B {\bf 494}, 471 (1997).

\bibitem{Bedaque-1998km}
  P.~F.~Bedaque, H.~W.~Hammer, and U.~van Kolck,
  Nucl.\ Phys.\ A {\bf 646}, 444 (1999).

\bibitem{Bedaque-1998kg}
  P.~F.~Bedaque, H.~W.~Hammer, and U.~van Kolck,
  Phys.\ Rev.\ Lett.\  {\bf 82}, 463 (1999).

\bibitem{Eden-Smatrix}
R.~J.~ Eden, P.~V.~Landshoff, D.~I.~Olive, and J.~C.~Polkinghorne,
{\it The Analytic $S$-matrix} (Cambridge University Press, 1966).

\bibitem{Bedaque:2000ft}
  P.~F.~Bedaque, E.~Braaten, and H.~W.~Hammer,
  Phys.\ Rev.\ Lett.\  {\bf 85} (2000) 908.


\bibitem{Barford:2002ut}
  T.~Barford and M.~C.~Birse,
  Few Body Syst.\ Suppl.\  {\bf 14} (2003) 123.


\bibitem{Platter:2006ev}
  L.~Platter and D.~R.~Phillips,
  Few Body Syst.\  {\bf 40} (2006) 35.

\bibitem{Pearce-1984ca}
  B.~C.~Pearce and I.~R.~Afnan,
  Phys.\ Rev.\ C {\bf 30}, 2022 (1984).


\bibitem{Helfrich-2010yr}
  K.~Helfrich, H.-W.~Hammer, and D.~S.~Petrov,
  Phys.\ Rev.\ A {\bf 81}, 042715 (2010).


\end{thebibliography}
\end{document}